\documentclass{aa}  
\usepackage{times,txfonts,latexsym,graphics,graphicx} 
   
\usepackage{natbib}
\bibpunct{(}{)}{;}{a}{}{,}

\newcommand{\teff}{$T_{\rm eff}$}
\newcommand{\logg}{$\log g$}
\newcommand{\vmic}{$\xi_{\rm t}$}
\newcommand{\vmac}{$\zeta_{\rm t}$}

\begin{document}    
  
\title{M-dwarf metallicities -- A high-resolution spectroscopic study in the near infrared
        \thanks{Based on data obtained at ESO-VLT, Paranal
        Observatory, Chile, Program ID 082.D-0838(A) and 084.D-1042(A).}} 
 
\author{Anna \"Onehag\inst{1}
\and    Ulrike Heiter\inst{1}
\and    Bengt Gustafsson\inst{1}
\and    Nikolai Piskunov\inst{1}
\and    Bertrand Plez\inst{2}
\and    Ansgar Reiners\inst{3}
}

\institute{Department of Physics and Astronomy, 
           Uppsala University,
           Box 516, S-751\,20 Uppsala, Sweden
\and Laboratoire Univers et Particules de Montpellier, CNRS, Universit\'e Montpellier
     2, 34095 Montpellier, France
\and       Universit\"at G\"ottingen, Institut f\"or Astrophysik, Friedrich-Hund-Platz 1, 
           37077 G\"ottingen, Germany}

\date{Recieved 15 September 2011 / Accepted 30 November 2011}  
\offprints{Anna \"Onehag,
\email{Anna.Onehag@fysast.uu.se}}

\authorrunning{\"Onehag et al.}
\titlerunning{M-dwarf metallicities}  

\abstract
{The relativley large spread in the derived metallicities ([Fe/H]) of M dwarfs shows that
 various approaches have not yet converged to consistency. 
 The presence of strong molecular features, and incomplete line lists 
 for the corresponding molecules have made metallicity determinations of M dwarfs difficult.
 Furthermore, the faint M dwarfs require long exposure times for a signal-to-noise 
 ratio sufficient for a detailed spectroscopic abundance analysis. 
 }
{We present a high-resolution (R$\sim$50,000) spectroscopic study of a sample of
 eight single M dwarfs and three wide-binary systems observed in the infrared J-band. 
}
{
 The absence of large molecular contributions allow for a precise 
 continuum placement. We derive metallicities based on the best fit
 synthetic spectra to the observed spectra. To verify the accuracy of 
 the applied atmospheric models and test our synthetic spectrum
 approach, three binary systems with a K-dwarf primary and an M-dwarf
 companion were observed and analysed along with the single M dwarfs.
}
{We obtain a good agreement between the metallicities derived for the
 primaries and secondaries of our test binaries and thereby confirm the 
 reliability of our method of analysing M dwarfs.
 Our metallicities agree well with certain earlier determinations,
 and deviate from others.
}
{We conclude that spectroscopic abundance analysis in the J band
 is a reliable method for establishing the metallicity scale for
 M dwarfs. We recommend its application to a larger sample 
 covering lower as well as higher metallicities. Further prospects of
 the method include abundance determinations for individual elements.
}


\maketitle

\section{Introduction}
Although the M dwarfs constitute a large
fraction of the detectable baryonic matter,
we still lack a great deal of knowledge about 
our low-mass ($<0.6M_{\odot}$) hydrogen-burning neighbours.
Studies \citep{1998ASPC..134...28H,2003PASP..115..763C} suggest that 
as much as 70\% of all stars in the solar vicinity ($\sim$10\,pc) are
M dwarfs which makes these objects essential
when deriving quantities such as the initial mass function 
\citep[IMF,][]{1955ApJ...121..161S} and the present day mass function.
These frequently used functions are derived using the luminosity. 
The transformation to mass, based upon stellar evolution theories,
is sensitive to chemical composition. 
Moreover, a study of the possible time dependence of the IMF function 
for the low-mass part, needs good stellar metallicity criteria for M dwarfs. 
Thus, if we are to create a realistic model of 
the galactic evolution and present day status, detailed studies of these 
faint but numerous objects are of great interest.
The M dwarfs are needed in the understanding of main-sequence stellar 
evolution and to define a limit between stellar and substellar objects. 
Finally, a well defined metallicity scale for M-dwarfs is essential to
determine whether or not the general trend towards supersolar metallicities
among FGK-stars planet hosts \citep[e.g.][]{2005ApJ...622.1102F}
holds also for cooler objects.

Spectroscopic studies of M dwarfs at high resolution have proven to be a difficult task.
In the low-temperature regime occupied by these targets
(2000$\lesssim$T$_{\rm eff}\lesssim$4100\,K), the optical spectrum is covered 
by a forest of molecular lines, hiding or blending most of the atomic lines used in 
spectral analysis.
However, models of low-mass late-type stars have undergone continuous improvements,  
from the early work by \citet{1969lls..symp..457T}, 
\citet{1969ApJ...157..799A} and \citet{1975A&A....38..283M,1976A&A....48..443M}, to the work by \citet{1993PASAu..10..250B},  
\citet{1995A&A...295..736B,1995A&AS..109..263B}, \citet{1995ApJ...445..433A}, 
\citet{1999ApJ...512..377H} with their extensive grid of NextGen models, and the 
improved MARCS models \citep{2008A&A...486..951G}. New molecular
and atomic line data are collected and organised in large databases such 
as VALD \citep{1999A&AS..138..119K} and VAMDC \citep{2010JQSRT.111.2151D} and
will improve the situation further.

The faintness characterising the M dwarfs has limited the number of 
high-resolution, high signal-to-noise studies. 
Non-sufficient resolution and dominant molecular features make it 
difficult to derive accurate atomic line strengths needed for a reliable metallicity
determination. 
Metallicities have been derived via photometric calibrations \citep{2005A_A...442..635B}, 
studies using molecular indices \citep{2006PASP..118..218W}, as well as spectrum synthesis 
\citep{2006ApJ...653L..65B,2006ApJ...652.1604B} and spectroscopic calibrations in the K band
\citep{2010ApJ...720L.113R}.

In this paper we present a detailed spectroscopic study in the J-band ($1100-1400$\,nm),
a spectral region relatively free from molecular lines. In the near infrared, atomic
lines can be isolated and the lack of molecular lines allows a precise continuum placement.
This spectral region was exploited for abundance analysis in the pioneering study of Betelgeuse,
based on FTS spectra, by \citet{VieiraThesis}, but has not been used much for  
late-type dwarfs until the last decade due to
lack of efficient IR spectrometers at large telescopes. The present  
generation of IR echelles at large telescopes have, however,
opened new possibilities, see, e.g. 
\citet{2000ApJ...533L..45M,2007ApJ...658.1217M}.

Similarly to previous studies \citep{2006ApJ...652.1604B,2005A_A...442..635B}, we select
a number of binary systems with a solar-type primary and an M-dwarf 
companion to assess the accuracy of the atmospheric models used in the analysis
and to verify the atomic line treatment.
We then apply this result to a number of non-binary M dwarfs.
Our choice of spectral region makes a careful analysis possible,
for a sample of stars that are thought to have a high metal content 
($-$0.35 to 0.5~dex), avoiding the large contribution of molecules such as TiO
to the spectrum at optical wavelengths.

The paper is organized as follows. In Section~\ref{sec:prevstud} we 
briefly summarize previous metallicity studies of M dwarfs.
In Section~\ref{sect:observations}, we describe the 
programme stars, the observations, and some aspects of the data reduction.
In Section~\ref{sect:analysis}, the ingredients of the spectrum analysis are 
presented -- compilation and derivation of spectral line data, stellar atmospheric 
parameters, and the procedure for metallicity estimation. 
In Section~\ref{sect:results} we discuss the results of the analysis, and 
Section~\ref{sect:conclusions} concludes the paper.

\section{Previous studies}
\label{sec:prevstud}
The stars in our sample have been investigated with various methods 
different from ours by several authors. In the following, we give a 
brief description of these studies.

\citet{2005A_A...442..635B} developed a photometric calibration of M-dwarf 
metallicities based on the spectroscopic analysis of F/G/K-type components 
of wide binary systems, where the secondaries are M dwarfs. They complemented 
their calibration sample with spectroscopic metallicities derived for 
metal-poor early-M-type dwarfs by \citet{2005MNRAS.356..963W}. Based on 
these metallicities and 2MASS photometry for 46 stars, they derived an 
expression for the metallicity as a function of absolute K magnitude $M_K$ 
and $V-K$ colour.

\citet{2006ApJ...653L..65B} analysed optical spectra with $R\ge50000$ 
and signal-to-noise ratios between 200 and 400 of three stars. They used 
the methods developed in \citet{2006ApJ...652.1604B}, fitting synthetic 
spectra for 16 atomic lines in the spectral intervals 8326 to 8427 and 8660 
to 8693~\AA, as well as a TiO bandhead at 7088~\AA\ to their observations. 
They simultaneously determined \teff, metallicity, broadening parameters, 
and continuum normalization factors from the spectra. \citet{2006ApJ...652.1604B} 
used five wide binary stars with F/G/K primaries and M-dwarf secondaries to 
evaluate their method (amongst them GJ~105). They found differences in derived 
metallicity between the M dwarf and solar-similar components ranging from $-0.16$ 
to $-0.07$ for four systems, and +0.03~dex for GJ~105.

\citet{2009ApJ...699..933J} and \citet{2010A&A...519A.105S} both aimed to 
improve the \citet{2005A_A...442..635B} photometric calibration, taking a 
slightly different approach.
First, they used a volume-limited calibration sample of solar-type stars to derive the 
mean metallicity of the solar neighbourhood. 
\citet{2009ApJ...699..933J} selected 109 G0-K2 dwarfs with spectroscopically 
determined metallicities and distances $d<18$~pc from \citet{2005ApJS..159..141V}.
\citet{2010A&A...519A.105S} selected a sample of F and G dwarfs with metallicity 
estimates based on Str\"omgren photometry and $d<20$~pc from \citet{2009A&A...501..941H}, 
which was kinematically matched to the solar-neighbourhood M-dwarf population.

Next, these authors defined a main-sequence line in the $(V-K) - M_K$ plane for a 
second calibration sample of late-K and M-type dwarfs. \citet{2009ApJ...699..933J} defined 
the second calibration sample of nearby low-mass stars to be a volume-limited sample 
of single K-type dwarfs ($d<20$~pc) and single M-type dwarfs ($d<10$~pc) based on 
parallaxes from Hipparcos and other sources. They fit a fifth-degree polynomial to 
the $V-K$ colours and $M_K$ magnitudes of these stars. \citet{2010A&A...519A.105S} 
adopted the main-sequence line of \citet{2009ApJ...699..933J} for their study.

A third calibration sample was used to find the variation of metallicity with horizontal 
or vertical distance from the main-sequence line ($\Delta(V-K)$ or $\Delta M_K$, respectively). 
\citet{2009ApJ...699..933J} used a set of six M dwarfs with FGK-companions with 
metallicities $>+0.2$~dex from \citet{2005ApJS..159..141V} and assigned to the main-sequence 
line the mean metallicity of the first calibration sample. They derived a linear relationship 
between [Fe/H] and $\Delta M_K$ with a dispersion of 0.06~dex. 
\citet{2010A&A...519A.105S} extended this calibration set by adding 13 wide-binary stars 
with accurate $V$ magnitudes from \citet{2005A_A...442..635B} with $-0.33\le$ [Fe/H] $\le+0.32$.
They derived a linear relationship between [Fe/H] and $\Delta(V-K)$, based only on the third
calibration sample and the main-sequence line. In this case, the first calibration sample was used
to verify that the zero-point of this relationship is close to the mean metallicity of the 
solar neighbourhood.

The work of \citet{2010ApJ...720L.113R} is based on low-resolution spectroscopy in the K-band. 
They used a calibration sample of 17 M dwarfs in wide-binary systems with metallicities 
determined for the FGK primaries by \citet{2005ApJS..159..141V}. From these metallicities 
and their observations, they derived a linear relationship between [Fe/H], two 
metallicity-sensitive indices measured from Na~I and Ca~I features, and a temperature-sensitive 
water index. They estimate an uncertainty for their calibration of 0.15~dex.

\begin{table*}[htbp]
  \caption{The target list. Period gives the ESO period in which our observations were taken. The
           references of the listed spectral types are: 1: \citet{2006AJ....132..161G}, 2: \citet{2007MNRAS.378.1328M}, 3:
           \citet{2009ApJ...704..975J}, 4: \citet{1972AJ.....77..486U}, 5:
           \citet{1995AJ....110.1838R}, 6: \citet{1996AJ....112.2799H}. The references of the planet 
           detections are: a: \citet{2005A&A...437.1121L}, b: \citet{2008ApJ...673.1165E}, c: 
           \citet{2007ApJ...670..833J}, d: \citet{2004ApJ...617..580B}, e: \citet{2005A&A...443L..15B},
           f: \citet{2007A&A...474..293B}, g: \citet{2006PASP..118.1685B}, h: \citet{1998ApJ...505L.147M}.}
  \centering
  \begin{tabular}[h]{llcccccccccc}
    Target & Other ID & Type &  Planet? & Period & $\sim$S/N      &  Ref.  &  Ref. Planet  \\ 
    \hline \hline
     HD 101930A   & HIP 57172  & K2        & yes & 82    & 120    &  1     &   a \\
     HD 101930B   & TYC 8638-366-1 & M0-M1 &     & 82    & 150    &  2     &     \\
     GJ 105 A     & HIP 12114  & K3        & no  & 82    & 110    &  1     &     \\
     GJ 105 B     & BD+06 398B & M4.5      &     & 84    & 80     &  3     &     \\
     GJ 250 A     & HIP 32984  & K3        & no  & 82    & 80     &  4     &     \\
     GJ 250 B     & HD 50281B  & M2        &     & 82    & 100    &  5     &     \\
     GJ 176       & HIP 21932  & M2        & yes & 82    & 70     &  5     &   b \\ 
     GJ 317       & LHS 2037   & M3.5      & yes & 84    & 100    &  5     &   c \\
     GJ 436       & HIP 57087  & M2.5      & yes & 84    & 140    &  5     &   d \\
     GJ 581       & HIP 74995  & M3        & yes & 84    & 110    &  5     &   e \\
     GJ 628       & HIP 80824  & M3.5      & no  & 84    & 130    &  5     &     \\
     GJ 674       & HIP 85523  & M2.5      & yes & 84    & 140    &  6     &   f \\
     GJ 849       & HIP 109388 & M3.5      & yes & 82,84 & 90,120 &  3     &   g \\
     GJ 876       & HIP 113020 & M4        & yes & 84    & 100    &  5     &   h \\
    \hline
   \end{tabular} 
   \label{tab:targets}
\end{table*}

\section{Target selection and observations}
\label{sect:observations}
We compiled a sample of M dwarfs in binary systems
with a solar-type (FGK) primary companion and non-binary M dwarfs
in the solar vicinity. Some of the M dwarfs or systems we observed are 
known to harbour planets, others have no detection 
of any planet companions as yet.
The programme stars were selected from the Catalogue of nearby wide 
binary and multiple systems \citep{1994RMxAA..28...43P} and from the Interactive 
Catalog of the on-line Extrasolar Planets Encyclopaedia 
\citep{2011A&A...532A..79S}\footnote{http://exoplanet.eu}, as well as from a 
programme searching for stellar companions of exoplanet host stars 
\citep{2004A&A...425..249M,2005A&A...440.1051M,2007MNRAS.378.1328M}.

\begin{figure*}[htbp]
  \centering
   \resizebox{\hsize}{!}{\includegraphics{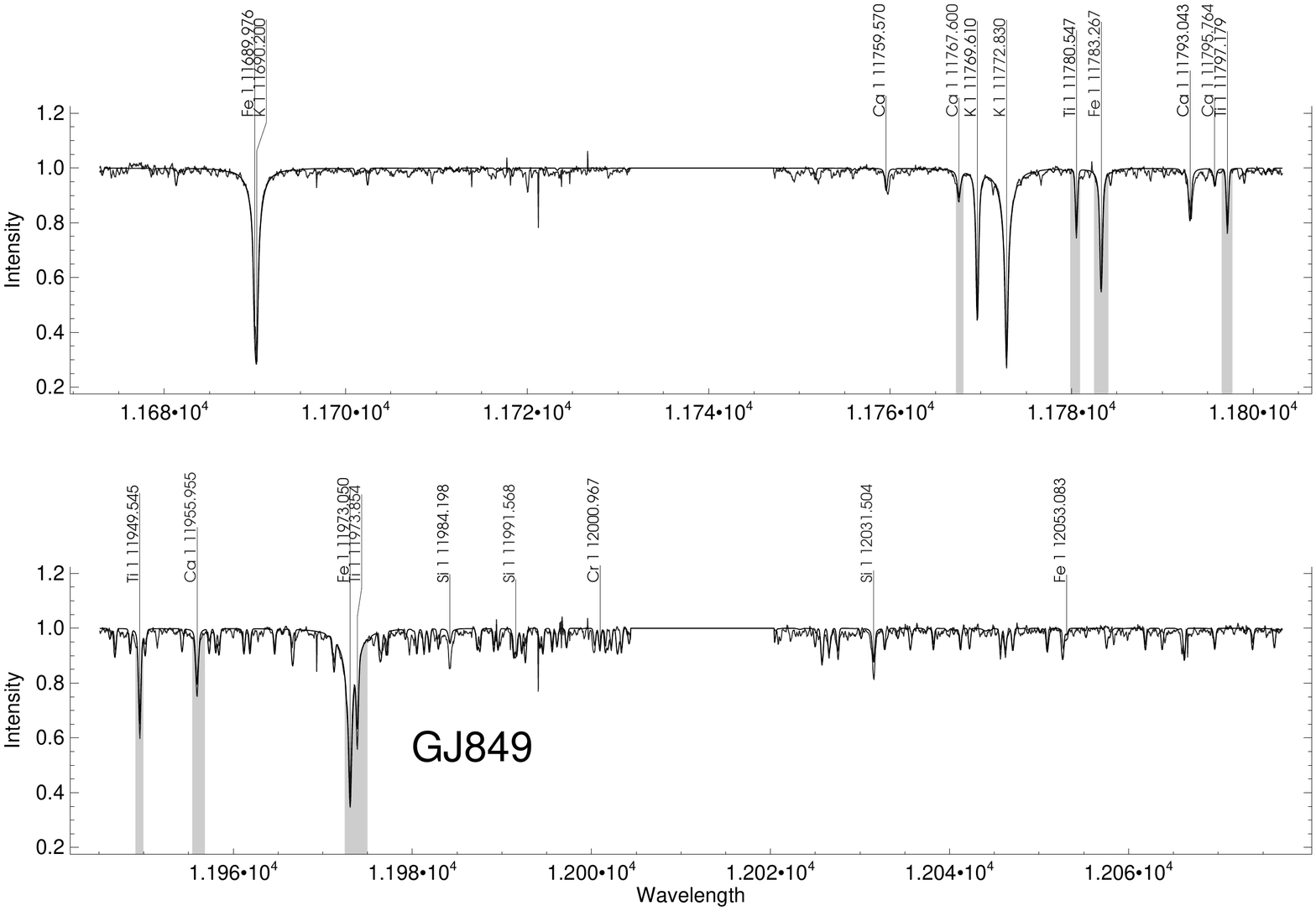}}
   \resizebox{\hsize}{!}{\includegraphics{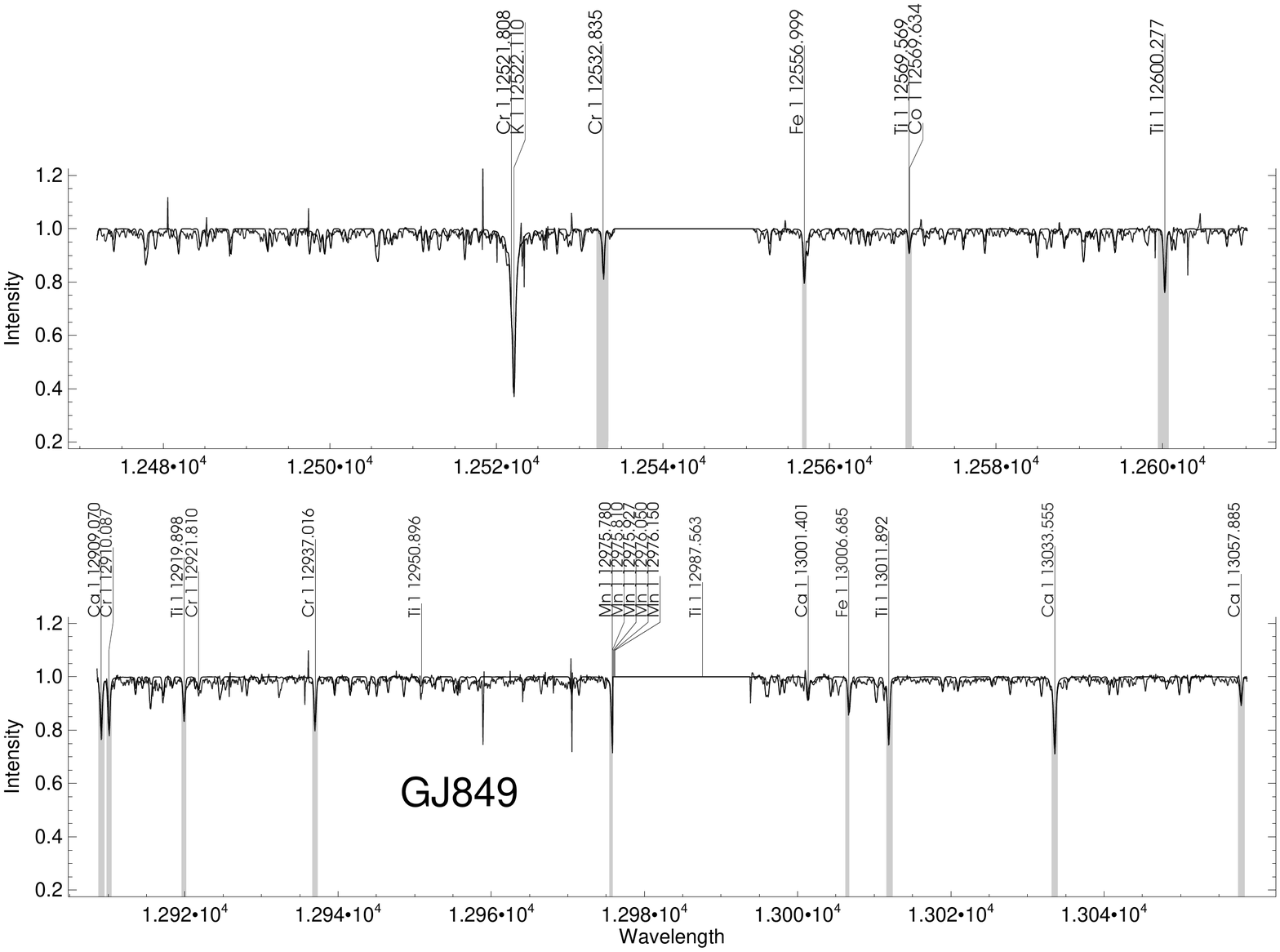}}
  \caption{The eight observed wavelength regions plotted together with 
           the best fit synthetic spectrum for one of the coolest targets, GJ849 
           observed in period 84. The mask used in the metallicity determination is marked
           with a grey shade.}
  \label{fig:GJ849}
\end{figure*}

The observations were carried out in service mode
with the infrared spectrometer CRIRES at ESO-VLT \citep{2004SPIE.5492.1218K}. 
In total 14 targets were observed during periods 82 (1st of October 2008 to 31st of March 2009) 
and 84 (1st of October 2009 to 31st of March 2010). 
A slit width of 0.4$\arcsec$ was used, 
resulting in a resolving power of R = $\lambda/\Delta\lambda = 50\,000$.
In addition a number of close binary systems with small separations ($\leq$\,20$\arcsec$) 
were observed which will be discussed in a future paper.
In this article we present the analysis of three wide binary systems and
eight single M dwarfs. The binary systems are well separated and
the angular separations are 73$\arcsec$ for HD~101930 \citep{2007MNRAS.378.1328M}, 165$\arcsec$ 
for GJ~105 \citep{1938ApJ....88...27V}, and 58.3$\arcsec$ for GJ~250 \citep{2002yCat.1274....0D}.
The observations of our targets should therefore not be contaminated with light 
from the companion star.
GJ~105A has a faint, close-by (3$\arcsec$) low-mass companion, GJ~105C 
\citep{1995ApJ...444L.101G,1995ApJ...452L.125G}. The luminosity difference
in the J band however is on the order of five magnitudes and the fainter companion
is assumed not to affect the analysis.
See Table~\ref{tab:targets} for a list of spectral types, binarity and planet detections of the
stars treated in this paper.
Each target was observed with four different CRIRES wavelength 
settings, centered on 1177, 1181, 1204, and 1258\,nm in period 82, and 1177, 1205, 1258, and
1303\,nm in period 84 (see Figure \ref{fig:GJ250A} \& \ref{fig:GJ849} for the total wavelength 
coverage). For some of the fainter targets we obtained several exposures, 
which were co-added to reach a signal-to-noise ratio around 100. 
The typical continuum signal-to-noise ratio spans between 70 and 150. 

CRIRES contains four detectors, but unfortunately
only detectors \#2 and \#3 produced reliable data, as \#1 and \#4 are
heavily vignetted and possibly contaminated by crosstalk
between adjacent orders. Realizing the extent of this failure of the first and fourth
detector we chose to rearrange the wavelength settings between the
observing periods. We re-observed one target from period 82 (GJ~849) in period 84 to assure
consistency between the two observing runs.
As is shown below (Section~\ref{sect:results}), our analysis indeed
gives the same metallicity for both periods, which supports the homogeneity of our observations. 
The analysis in this paper is based on the higher reliability data from detectors 
\#2 and \#3.

In connection with each observation a rapidly rotating early-type star, was observed to 
represent the telluric spectrum.
Although the observed region (1167--1306\,nm) was chosen to harbour as few
telluric lines as possible, the majority of lines detected in the 
spectra still were of telluric origin. The pipe-line reduced spectra
were normalised together with the corresponding telluric standard to ensure
a consistent continuum placement. The absence of strong molecular 
absorptions made continuum windows easily recognizable.\\
\indent
From a first examination of the reduced data it became clear that the wavelength 
calibration in ESO's reduction pipe-line
based on thorium and argon lines did not produce the desired outcome. 
Both overall shifts and distortions in the wavelength scale could be seen,
compared to the solar atlas or to synthetic spectra, probably because of the 
small number of thorium and argon lines present in the calibration frame of the 
observed wavelength regions.
The solution was to make use of the telluric lines present in the observations
and to match these with telluric lines in the electronic version of the atlas 
of the solar spectrum 
\citep{1991aass.book.....L}\footnote{ftp://nsokp.nso.edu/pub/Kurucz\_1984\_atlas/photatl/}, 
using a polynomial fit.

\section{Analysis}
\label{sect:analysis}

\subsection{Spectral line data}
The atomic line data in the observed region were acquired from the 
Vienna Atomic Line Database \citep[VALD][]{1999A&AS..138..119K},
with the exception of a few lines that are from \citet{1999ApJS..124..527M}. 
We used the Sun as a reference and calculated a synthetic
spectrum using the established line list. A MARCS model with the parameters
\teff= 5777\,K, \logg= 4.44, [Fe/H]= 0.00 was adopted and $v\sin i$= 0.7\,kms$^{-1}$ was used.
We used the same solar chemical composition as in the MARCS models \citep{2007SSRv..130..105G}.
The unknown line-broadening micro- (\vmic) and macroturbulence (\vmac) parameters,
must be adjusted when comparing the synthetic spectrum with observed
lines. We used a high-quality solar spectrum where telluric lines
have been removed \citep{1991aass.book.....L} and the SME package 
(see Section~\ref{sect:sme}) to solve for both turbulence parameters.

When comparing our solution with the observed spectrum we noted that
a few lines did not match. This might be the result of inaccuracies in the listed 
oscillator strengths and damping parameters and we therefore determined 
new $\log gf$ and van der Waals broadening parameters for these 
particular lines (assuming that hydrogen is the main perturber in this temperature regime) . 
This was done in an iterative scheme where 
we first solved for the $\log gf$ and van der Waals parameters for each of the deviating 
lines separately, and then determined the turbulence parameters using all lines.
After a few iterations we converged to a synthetic fit that
reproduced the solar line profiles. The \vmic\ and \vmac\ values that
yielded the best fit were found to be 0.79\,kms$^{-1}$ and 1.77\,kms$^{-1}$, respectively.
The final line data for the dominant lines used in the metallicity analysis 
can be found in Table~\ref{tab:lines}, where we have marked the lines for which 
we re-determined the line parameters.
  
The observed wavelength region was chosen to contain as few stellar molecular 
features as possible. Significant stellar molecular absorption in the observed 
wavelength regions comes from FeH.  In addition, some absorption from CrH and 
water is expected.

Spectral lines of FeH were synthesized for all targets together with the atomic lines, 
to account for possible blends.  The FeH line list was calculated by one of us (BP), 
using the best available laboratory data for energy levels and transition moment 
\citep{1987ApJS...65..721P,1990JMoSp.141..243L}. The weak FeH lines visible in the spectra of 
our M-dwarf targets seem to be reproduced rather well.
We compared our FeH line list with that of \citet{2003ApJ...594..651D} by calculating 
spectra with both lists for the two stars GJ~436 and GJ~628.
An overall comparison showed that the FeH lines calculated with the 
\citet{2003ApJ...594..651D} list were somewhat weaker than when using the BP list.
Line-depth ratios between a number of selected FeH lines appearing in both lists and in 
the observations were calculated. For this calculation, we used mean fluxes of the 
40\% of the pixels closest in wavelength to the line center within the regions masking 
each FeH line, such as those shown in Fig.~\ref{fig:fehteff}.
The BP ratios were closer to the observed ones than the Dulick ratios for a majority of 
these lines -- for GJ~436 for 20 out of 30 lines, and for GJ~628 for 22 out of 37 lines.
The spectra calculated with the \citet{2003ApJ...594..651D} list contain several lines 
which do not appear in the BP list, and which we do not observe in our spectra.
In conclusion, we decided to use the BP list for the analysis.
After the completion of the present paper we noted that the line-list by 
\citet{2003ApJ...594..651D} has been used by \citet{2010A&A...523A..58W} to model
CRIRES spectra of a late M dwarf in the wavelength range 986--1077\,nm and subsequently 
by \citet{2011arXiv1108.3465S} for a study of rotation and magnetic fields
in late-type M-dwarf binaries.

We also synthesized the observed spectral regions including CrH lines with data taken 
from \citet{2002ApJ...577..986B} for a representative set of parameters. 
The regions contain a few weak CrH lines, but they do not coincide with any of the atomic 
lines selected for analysis, and thus were not taken into account.

We assessed the importance of water absorption for our spectral region by computing 
synthetic spectra using the line list of \citet{2006MNRAS.368.1087B}.
From the $\approx$27 million theoretical transitions listed between 1160 and 1320~nm, 
we removed those with a line strength of less than 0.5\% of the strongest line at 
$T$=3000~K (the line strength measure was $\log(\lambda)+\log(gf)-E_{\rm low}/(\ln10kT)$), 
resulting in 57265 lines.
We computed pure water spectra for atmospheric models with a range in \teff\ and 
metallicity corresponding to our M-dwarf sample. 
If the line parameters are correct, spectra with \teff=3200~K may suffer from a decrease 
in the continuum level of up to 2\%, caused by numerous weak water lines. For higher 
temperatures, the importance of water absorption decreases rapidly.
For wavelengths less than about 1200~nm, individual water lines with depths of up to 5\% 
are apparent in the test calculations for \teff=3200~K.
We also calculated spectra for the parameters of GJ~628 (see Section~\ref{sect:atmparam}) 
for the four wavelength segments with $\lambda <$ 1208~nm, including atomic, FeH, and water 
lines, and compared them with the observations. 
There was a certain resemblance between some of the synthetic water lines and some of the otherwise 
unidentified features in the observed spectra, but we could not verify the accuracy of the 
wavelengths and line strengths to a satisfactory degree. Also, the profiles of the atomic 
lines showed little change in the spectra with and without water absorption. Hence, we decided 
not to include the water line list in the analysis.

\begin{table*}[htbp]
   \caption{Atomic lines used in analysis. $E_{\rm low}$ is the lower level energy. Column ``VdW'' 
    lists the parameter(s) used to calculate line broadening due to van der Waals interaction. 
    Negative numbers give the logarithm of the line width per perturber at l0$^4$~K in rad s$^{-1}$ 
    cm$^3$, log$\gamma_{\rm W}$. For positive numbers, the integer part gives the broadening 
    cross-section at a velocity of 10$^4$~m s$^{-1}$ in atomic units, while the fractional part 
    gives the velocity parameter \citep[see][]{BPM}. Column ``Source'' gives a reference code for 
    the atomic data (see notes below table), or an ``S'', meaning that log$gf$ and log$\gamma_{\rm W}$ 
    were determined from a fit to the solar spectrum, with the reference code in parentheses 
    indicating the source of the initial data. When there are two codes, the first one refers 
    to log$gf$, and the second one to log$\gamma_{\rm W}$. K or M in the last column indicates 
    that the line was used in the K or M dwarf analysis, respectively (some of them only in the 
    period 82 (P82) or period 84 (P84) spectra, see Table~\ref{tab:targets}).}
  \centering
  \begin{tabular}{llrrrcc}
     Wavelength\,[\AA] & Species & $E_{\rm low}$ [eV] & log$gf$& VdW & Source & K/M \\       
    \hline \hline                                                                            
    11681.594         & Fe I    & 3.547 & $-$3.301 &  $-$7.352 & S(K07)      & K   \\    
    11682.250         & Fe I    & 5.620 & $-$1.420 &  $-$7.520 & MB99        & K   \\    
    11715.487         & Fe I    & 5.642 & $-$0.961 &  $-$7.142 & S(K07)      & K   \\    
    11725.563         & Fe I    & 5.699 & $-$1.279 &  $-$7.112 & S(K07)      & K   \\    
    11727.733         & Fe I    & 6.325 & $-$0.879 &  $-$6.722 & S(K07)      & K   \\    
    11743.695         & Fe I    & 5.947 & $-$0.943 &  $-$6.932 & S(K07)      & K (P82) \\
    11767.600         & Ca I    & 4.532 & $-$0.635 &  $-$6.777 & S(K07)      & K,M \\    
    11780.547         & Ti I    & 1.443 & $-$2.180 &  $-$7.790 & BLNP,K10    & K,M \\    
    11783.267         & Fe I    & 2.832 & $-$1.520 &  $-$7.842 & S(BWL,K07)  & K,M \\    
    11783.433         & Mn I    & 5.133 & $-$0.094 &  $-$7.560 & K07         & K \\      
    11797.179         & Ti I    & 1.430 & $-$2.250 &  $-$7.790 & BLNP,K10    & K,M \\    
    11828.171         & Mg I    & 4.346 & $-$0.046 &   862.225 & S(N10),BPM  & K,M (P82) \\
    11949.545         & Ti I    & 1.443 & $-$1.550 &  $-$7.790 & BLNP,K10    & K,M \\     
    11949.760         & Ca II   & 6.470 & $-$0.040 &           & MB99        & K   \\     
    11955.955         & Ca I    & 4.131 & $-$0.849 &  $-$7.300 & K07         & K,M \\     
    11973.050         & Fe I    & 2.176 & $-$1.405 &  $-$7.889 & S(BWL,K07)  & K,M \\     
    11973.854         & Ti I    & 1.460 & $-$1.591 &  $-$7.528 & S(BLNP,K10) & K,M \\     
    11984.198         & Si I    & 4.930 & $+$0.239 &   677.228 & K07,BPM     & K   \\     
    11991.568         & Si I    & 4.920 & $-$0.109 &   674.228 & K07,BPM     & K   \\     
    12031.504         & Si I    & 4.954 & $+$0.477 &   685.229 & K07,BPM     & K   \\          
    12039.822         & Mg I    & 5.753 & $-$1.530 &           & N10         & K   \\     
    12044.055         & Cr I    & 3.422 & $-$1.863 &  $-$6.281 & K10         & K   \\     
    12044.129         & Fe I    & 4.988 & $-$2.130 &  $-$6.677 & K07         & K   \\     
    12053.083         & Fe I    & 4.559 & $-$1.543 &  $-$7.540 & BWL,K07     & K   \\         
    12510.520         & Fe I    & 4.956 & $-$1.846 &  $-$7.142 & S(K07)      & K   \\         
    12532.835         & Cr I    & 2.709 & $-$1.879 &  $-$7.800 & K10         & K,M \\         
    12556.999         & Fe I    & 2.279 & $-$3.913 &  $-$7.422 & S(BWL,K07)  & K,M \\            
    12569.569         & Ti I    & 2.175 & $-$1.867 &  $-$7.810 & K20         & K,M \\             
    12569.634         & Co I    & 3.409 & $-$0.992 &  $-$7.730 & K21         & K   \\             
    12600.277         & Ti I    & 1.443 & $-$2.150 &  $-$7.790 & BLNP,K10    & K,M \\     
    12909.070         & Ca I    & 4.430 & $-$0.426 &  $-$7.787 & K20         & M (P84) \\    
    12910.087         & Cr I    & 2.708 & $-$1.863 &  $-$7.402 & S(K10)      & M (P84) \\ 
    12919.898         & Ti I    & 2.154 & $-$1.553 &  $-$7.750 & K10         & M (P84) \\ 
    12937.016         & Cr I    & 2.710 & $-$1.896 &  $-$7.800 & K10         & M (P84) \\ 
    12975.927         & Mn I    & 2.888 & $-$1.356 &  $-$7.372 & S(K07)      & M (P84) \\ 
    12975.72-12976.15$^a$ & Mn I &       &        &        & S(MB99)     & M (P84) \\     
    13001.401         & Ca I    & 4.441 & $-$1.139 &  $-$7.710 & K07         & M (P84) \\  
    13006.685         & Fe I    & 2.990 & $-$3.269 &  $-$7.412 & S(K07)      & M (P84) \\ 
    13011.892         & Ti I    & 1.443 & $-$2.180 &  $-$7.790 & BLNP,K10    & M (P84) \\ 
    13033.555         & Ca I    & 4.441 & $-$0.064 &  $-$7.710 & K07         & M (P84) \\ 
    13057.885         & Ca I    & 4.441 & $-$1.092 &  $-$7.710 & K07         & M (P84) \\ 
    \hline
   \end{tabular}
   \tablefoot{$^a$ Blend of eight Mn~I lines. References: BLNP -- \citet{BLNP}, BPM -- \citet{BPM}, 
              BWL -- \citet{BWL}, K20 -- \citet{1994KurCD..20.....K}, K21 -- \citet{1994KurCD..21.....K}, 
              K07 -- \citet{K07}, K10 -- \citet{K10}, MB99 -- \citet{1999ApJS..124..527M}, 
              N10 -- \citet{NIST10,6692EL,5539TP}.}
   \label{tab:lines}
\end{table*}

\subsection{Atmospheric parameters}
\label{sect:atmparam}

\begin{table*}[htbp]
   \caption{Compiled atmospheric parameters for the K stars. References: 1: \citet{2008A_A...487..373S}, 2: \citet{2005A_A...437.1127S}
    3: \citet{2004A&A...415.1153S}, 4: \citet{2004A&A...418..551M}, 5: \citet{2005ApJS..159..141V}, 6: \citet{2006AJ....131.3069L},
       7: \citet{2003AJ....126.2015H}, 8: \citet{2006ApJ...652.1604B}, 9: \citet{2005A&A...437.1121L}. The $v\sin i$ reference is stated to the right in
       the list of references. The miroturbulent velocity for GJ105A is from reference 6,7,8, and from references 3,6,7 for target GJ250A.}
  \centering
  \begin{tabular}{lccccc}
    Target         & \teff\,[K]   & \logg\,[cms$^{-2}$] & \vmic\,[kms$^{-1}$]     & $v\sin i$\,[kms$^{-1}$]  & References  \\ 
    \hline \hline
     HD101930 A    & 5121\,$\pm$\,87  & 4.32\,$\pm$\,0.19 & 0.82\,$\pm$\,0.14 & 0.7   & 1,2,       9 \\
     GJ 105 A      & 4867\,$\pm$\,114 & 4.60\,$\pm$\,0.07 & 0.60\,$\pm$\,0.32 & 2.9   & 5,6,7,8,   5 \\ 
     GJ 250 A      & 4758\,$\pm$\,173 & 4.40\,$\pm$\,0.33 & 0.64\,$\pm$\,0.56 & 1.8   & 3,4,5,6,7, 5 \\ 
     \hline
   \end{tabular} 
   \label{tab:paramK}
\end{table*}

\begin{table*}[htbp]
   \caption{Stellar colours. All $J, H$, and $K_{\rm S}$ values are from 2MASS \citep{2003tmc..book.....C}.
   References for $B, V, R_{\rm C}$, and $I_{\rm C}$:
   B90\,:\,\citet{1990A&AS...83..357B}, 
   C62\,:\,\citet{1962RGOB...64..103C}, 
   C80\,:\,\citet{1980SAAOC...1..166C}, 
   K01\,:\,\citet{2001KFNT...17..409K}, 
  Ki98\,:\,\citet{1998MNRAS.294...93K}, 
  Ki07\,:\,\citet{2007MNRAS.380.1261K}, 
   K02\,:\,\citet{2002MNRAS.334...20K}, 
   K10\,:\,\citet{2010MNRAS.403.1949K}, 
   L89\,:\,\citet{1989SAAOC..13...29L}, 
   R04\,:\,\citet{2004AJ....128..463R}, 
   T84\,:\,\citet{1984A&A...132..385T}  
   }
  \centering
  \begin{tabular}{lcccccccc}
     Target      &   $B$   &   $V$   &   $R_{\rm C}$   &  $I_{\rm C}$  &    $J$  & $H$  & $K_{\rm S}$ & Reference \\
     \hline \hline
     HD101930 A  &   9.12  &  8.21   &         &       &   6.645\,$\pm$\,0.019 &  6.259\,$\pm$\,0.047 &  6.147\,$\pm$\,0.026 &  C62 \\
     HD101930 B  &  11.663 &  10.605 &         &       &   7.940\,$\pm$\,0.025 &  7.291\,$\pm$\,0.049 &  7.107\,$\pm$\,0.024 &  K01 \\
     GJ 250 A    &   7.64  &  6.59   &  5.975  & 5.45  &   5.013\,$\pm$\,0.252 &  4.294\,$\pm$\,0.258 &  4.107\,$\pm$\,0.036 &  B90,C80 \\
     GJ 250 B    &  11.57  &  10.08  &  9.04   & 7.80  &   6.579\,$\pm$\,0.034 &  5.976\,$\pm$\,0.055 &  5.723\,$\pm$\,0.036 &  L89 \\ 
     GJ 105 A    &   6.78  &  5.81   &  5.235  & 4.74  &   4.152\,$\pm$\,0.264 &  3.657\,$\pm$\,0.244 &  3.481\,$\pm$\,0.208 &  B90,C80 \\
     GJ 105 B    &  13.16  &  11.66  &  10.44  & 8.88  &   7.333\,$\pm$\,0.018 &  6.793\,$\pm$\,0.038 &  6.574\,$\pm$\,0.020 &  L89 \\
     GJ 176      &  11.50  &  9.966  &  8.941  & 7.711 &   6.462\,$\pm$\,0.024 &  5.824\,$\pm$\,0.033 &  5.607\,$\pm$\,0.034 &  K10,R04 \\ 
     GJ 317      &  13.488 &  11.985 & 10.862  & 9.375 &   7.934\,$\pm$\,0.027 &  7.321\,$\pm$\,0.071 &  7.028\,$\pm$\,0.020 &  B90,L89 \\ 
     GJ 436      &  12.17  &  10.65  &  9.58   & 8.24  &   6.900\,$\pm$\,0.024 &  6.319\,$\pm$\,0.022 &  6.073\,$\pm$\,0.016 &  R04  \\ 
     GJ 581      &  12.179 &  10.571 &  9.456  & 8.051 &   6.706\,$\pm$\,0.025 &  6.095\,$\pm$\,0.033 &  5.837\,$\pm$\,0.022 &  B90,L89,K02,K10,T84 \\ 
     GJ 628 	 &  11.657 &  10.082 &  8.918  & 7.410 &   5.950\,$\pm$\,0.024 &  5.373\,$\pm$\,0.040 &  5.075\,$\pm$\,0.024 &  B90,L89,K02,Ki98,Ki07 \\ 
     GJ 674      &  10.944 &   9.389 &  8.312  & 6.979 &   5.711\,$\pm$\,0.019 &  5.154\,$\pm$\,0.033 &  4.855\,$\pm$\,0.018 &  B90,L89,K10,T84 \\ 
     GJ 849      &  11.878 &  10.376 &  9.284  & 7.877 &   6.510\,$\pm$\,0.024 &  5.899\,$\pm$\,0.044 &  5.594\,$\pm$\,0.017 &  B90,L89,K02,K10,T84 \\ 
     GJ 876      &  11.759 &  10.187 &  9.006  & 7.446 &   5.934\,$\pm$\,0.019 &  5.349\,$\pm$\,0.049 &  5.010\,$\pm$\,0.021 &  B90,K02,Ki98,Ki07,T84 \\  
     \hline
   \end{tabular} 
   \label{tab:colours}
\end{table*}

\begin{table*}[htbp]
   \caption{Atmospheric parameters for the M dwarfs. The adopted effective temperatures, based on the 
            photometric calibration by \citet{2008MNRAS.389..585C} and FeH spectrum adjustment are given 
            in the second column. For comparison, the effective temperatures derived from the 
            \citet[][WL11]{2011ApJS..193....1W} calibration are also shown. Parallaxes are from 
            Hipparcos \citep{2007ASSL..350.....V} for all targets except for GJ~317 \citep{2007ApJ...670..833J} and
            GJ105 B \citep{2009ApJ...704..975J}. The $v\sin i$ references listed in the rightmost column are:
            1: \citet{2010AJ....139..504B}, 2: \citet{1992ApJ...390..550M}, 3: \citet{2007A&A...467..259R}, 
            4: \citet{2006A&A...460..695T}, 5: \citet{1998A&A...331..581D}.}
  \centering
  \begin{tabular}{llcccrcc}
    Target & \teff\,[K]& \teff\,[K] (WL11) & \logg\,[cms$^{-2}$] & Mass [$M_{\odot}$] & Parallax [mas] & $v\sin i$\,[kms$^{-1}$] & Ref.\\ 
    \hline \hline
     HD101930 B    & \bf{3908}          &  3887    & 4.46\,$\pm$\,0.09 & 0.70\,$\pm$\,0.02 & 34.24\,$\pm$\,0.81  &                 &   \\
     GJ 105 B      & \bf{3261}$\dagger \dagger$ &  3162    & 4.96\,$\pm$\,0.08 & 0.26\,$\pm$\,0.01 & 129.4\,$\pm$\,4.3   & $\leq$\,2.4     & 1 \\
     GJ 250 B      & \bf{3376}          &  3462    & 4.80\,$\pm$\,0.08 & 0.44\,$\pm$\,0.01 & 114.81\,$\pm$\,0.44 & $\leq$\,2.5     & 1 \\
     GJ 176        & \bf{3361}          &  3462    & 4.76\,$\pm$\,0.08 & 0.49\,$\pm$\,0.02 & 107.83\,$\pm$\,2.85 &                 &   \\
     GJ 317        & \bf{3325}$\dagger \dagger$ &  3199    & 4.97\,$\pm$\,0.12 & 0.25\,$\pm$\,0.06 &    109\,$\pm$\,20   & $\leq$\,2.5     & 1 \\
     GJ 436        & \bf{3263}          &  3376    & 4.80\,$\pm$\,0.08 & 0.44\,$\pm$\,0.01 & 98.61 \,$\pm$\,2.33 & 1.0\,$\pm$0.9   & 2 \\
     GJ 581        & \bf{3308}$\dagger$ &  3288    & 4.92\,$\pm$\,0.08 & 0.30\,$\pm$\,0.01 & 160.91\,$\pm$\,2.61 & 0.4\,$\pm$0.3   & 2 \\
     GJ 628        & \bf{3208}$\dagger$ &  3167    & 4.93\,$\pm$\,0.08 & 0.29\,$\pm$\,0.01 & 232.98\,$\pm$\,1.60 & 1.5             & 3 \\
     GJ 674        & \bf{3305}          &  3376    & 4.88\,$\pm$\,0.08 & 0.35\,$\pm$\,0.01 & 220.24\,$\pm$\,1.42 & 3.2\,$\pm$\,1.2 & 4 \\
     GJ 849        & \bf{3196}          &  3258    & 4.76\,$\pm$\,0.08 & 0.49\,$\pm$\,0.01 & 109.94\,$\pm$\,2.07 & $\leq$\,2.4     & 5 \\
     GJ 876        & \bf{3156}$\dagger$ &  3167    & 4.89\,$\pm$\,0.08 & 0.33\,$\pm$\,0.01 & 213.28\,$\pm$\,2.12 & $\leq$\,2.0     & 5 \\
    \hline
   \end{tabular} 
   \tablefoot{$\dagger$ and $\dagger \dagger$ denotes an increase of the photometric temperature by +100 and +200\,K, respectively.}
   \label{tab:paramM}
\end{table*}

\begin{figure*}
  \centering
   \resizebox{\hsize}{!}{\includegraphics{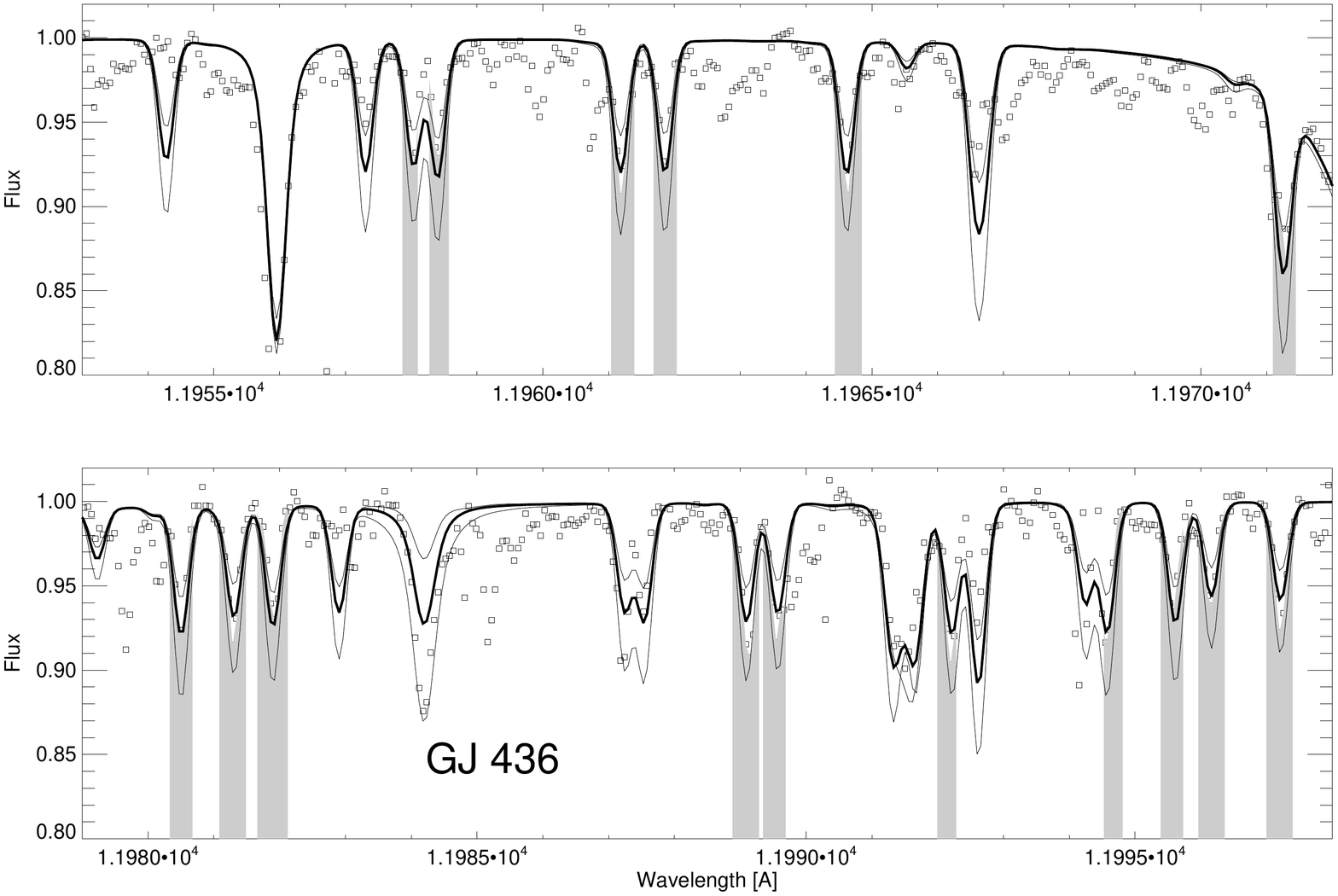}}
  \caption{Two selected wavelength regions containing FeH lines, with synthetic and observed spectra 
           for GJ~436. The FeH lines are marked with a grey shade. The thick solid line was calculated 
           using the adopted \teff of 3263\,K. The upper and lower thin lines are calculated with 200~K higher 
           and lower \teff\ values, respectively. The observations are represented by open squares.}
  \label{fig:fehteff}
\end{figure*}

The analysis using synthetic spectra requires a specification of several input parameters:
effective temperature, \teff, surface gravity, \logg, the macro- and microturbulence
parameters, and the overall metallicity [Fe/H] compared to the Sun.
We searched the archives (e.g. SIMBAD, VizieR) to find reliable
measures of the atmospheric parameters. We found spectroscopically
determined values for the three K-dwarf stars in our sample and used
an unweighted mean based on these values.  The adopted atmospheric parameters can 
be found in Table~\ref{tab:paramK}.
For the majority of the M-dwarf targets there are no detailed atmospheric 
studies available. To keep the analysis consistent, effective temperatures and 
surface gravities for the M-dwarf targets were determined using
calibration equations based on
photometric data. High-accuracy photometric data were 
available in the archives for all of our stars.
Photometric data in the Johnson $(BV)$ and Cousins $(RI)_{\rm c}$ systems
were collected, and where multiple measures were available an unweighted 
mean was calculated. Infrared colours ($JHK$) were gathered from the
Two Micron All Sky Survey \citep[2MASS:][]{2003tmc..book.....C},
see Table~\ref{tab:colours}.\\ 
\indent
Effective temperatures for the M dwarfs were estimated from
the photometric calibrations derived by \citet{2008MNRAS.389..585C}, using
different combinations of the $V$, $R_{\rm c}$, $I_{\rm c}$, $J$, $H$, $K_{\rm s}$ colours. 
The resultant twelve \teff\ values per target where then merged into a mean.
We noted that our main molecular features, the FeH lines, are quite temperature sensitive,
and used this sensitivity to verify and adjust the photometric temperatures.
The FeH lines show a good agreement for most photometric temperatures,
whereas for a few objects, the FeH lines indicate a correction by up to +200\,K 
(see Table \ref{tab:paramM}).
We estimated the uncertainties of the photometric temperatures by propagating the 
uncertainties of the observed colours through the calibrations. To this we added 
in quadrature the quoted uncertainty of the calibration expression itself. 
The resultant uncertainties of $\sim$150\,K for the photometric temperatures 
is in good agreement with the corrections we apply based on the temperature sensitivity of FeH.
An example is shown in Fig.~\ref{fig:fehteff}, where a synthetic spectrum 
with three different \teff\ values is compared to a region of the observed spectrum of GJ~436  
containing several FeH lines. The \teff\ values correspond to the photometric \teff, as well as 
200~K lower and higher values. We adjusted the temperature
and metallicity for four of the M-dwarf targets iteratively, since the FeH line strengths also depend
on the overall metallicity. We tested the sensitivity of the FeH lines to surface
gravity and did not find any significant effect in the parameter space and 
wavelength regions relevant to this study, contrary to \citet{2009A&A...508.1429W}
who find a surface gravity sensitivity using 3D-models. That study, however,
was carried out in a different wavelength region (997\,nm) and with a larger stepsize
in \logg\ than tested for here.

In addition, the tabulated grids of colours by \citet{2011ApJS..193....1W} were
used to estimate a value for the effective temperature, gravity, and metallicity 
simultaneously from an independent calibration.
The best-fit set of parameters was determined via a chi-square
minimization between observed and tabulated colours ($B-V, J-K, H-K, V-K, V-I, V-R$). This 
method resulted in \teff\ values similar to the adjusted \citet{2008MNRAS.389..585C}
calibrations (any differences are below 100~K). In Table~\ref{tab:paramM} we list effective 
temperatures derived from both methods.
We take the maximum difference between the two \teff\ determinations as an indication for
the range of \teff\ values to explore in the analysis for each target.

The surface gravities of the M dwarfs were established using the
\logg--${\cal M_{\rm \star}}$ relation derived 
by \citet[][their Eq.~2]{2006ApJ...652.1604B}. The input masses were
estimated from the $\log\,({\cal M}/{\cal M_\odot})- M_{K}$
relationship established by \citet{2000A&A...364..217D}. The $M_{K}$
in this relation refers to the absolute $K$ magnitude in the $CIT$ system
and a transformation of the 2MASS $K_{\rm S}$ magnitudes was carried out using
the equation presented in \citet{2001AJ....121.2851C}. Absolute magnitudes
in $K_{\rm CIT}$ were derived using the Hipparcos parallaxes \citep{2007ASSL..350.....V}
for all targets except for GJ~317 and GJ~105~B where the parallaxes were taken from \citet{2007ApJ...670..833J}
and \citet{2009ApJ...704..975J}, respectively.
An estimate of the total \logg\ error was made by propagating the uncertainties 
in the 2MASS $K_{\rm S}$ colours and listed parallax uncertainties through the 
calibration equations. In the absence of an error estimate of the mass-luminosity relation
we derived the standard deviation of the mass-luminosity fit using the 
data on which the calibration expression is based \citep[][their Table~3]{2000A&A...364..217D}.
The surface gravity--mass calibration is determined by \citet{2006ApJ...652.1604B}
to have an error of 0.08 [log(cms$^{-2}$)] in \logg.
Ignoring possible errors in the $K_{\rm S}$--$K_{\rm CIT}$ transformation,
the different uncertainties were added in quadrature to calculate the final
overall error (see Table~\ref{tab:paramM}).

\subsection{Model atmospheres and abundance determination}
\label{sect:sme}

\begin{table*}[htbp]
   \caption{Derived metallicities for our sample and comparison with previous 
            abundance determinations: Bo05: \citet{2005A_A...442..635B}, 
            Be06: \citet{2006ApJ...653L..65B}, JA09: \citet{2009ApJ...699..933J},
            SL10: \citet{2010A&A...519A.105S}, R10: \citet{2010ApJ...720L.113R}.            
            The [Fe/H] marked ``c'' in the JA09 column denotes that this star was
            not present in their study and we therefore used our collected colours and
            their provided calibration. Note that the metallicity determination of 
            HD101930B from JA09 is an extrapolation slightly outside the valid range.
            The metallicities in the Bo05 and SL10 columns are 
            calculated from their respective calibration equations except for
            the values marked with an "s" that are spectroscopically determined. 
            The errors quoted for this work are a combination of systematic errors
            from the uncertainties in atmospheric parameters and the uncertainty from 
            the fitting routine (see Section \ref{sect:sme}).}
  \centering
  \begin{tabular}{lllllll}
    Target        &  This work                 & Bo05      & Be06    & JA09       & SL10     &   R10    \\
    \hline \hline                                                                               
    HD101930A     & {\bf $+$0.20\,$\pm$\,0.06} &           &         &            &          &          \\   
    HD101930B     & {\bf $+$0.09\,$\pm$\,0.10} & $-$0.16   &         & $+$0.08\,c & $-$0.05  &          \\   
    GJ105A        & {\bf $-$0.05\,$\pm$\,0.01} & $-$0.19\,s& $-$0.12 &            &          &          \\   
    GJ105B        & {\bf $-$0.06\,$\pm$\,0.15} & $-$0.15   & $-$0.09 & $+$0.06\,c & $-$0.07  &  $-$0.04 \\   
    GJ250A        & {\bf $-$0.03\,$\pm$\,0.05} & $-$0.15\,s&         &            &          &          \\   
    GJ250B        & {\bf $-$0.05\,$\pm$\,0.05} & $-$0.18   &         & $+$0.05\,c & $-$0.09  &          \\   
    \hline                                                                   
    GJ176         & {\bf $+$0.04\,$\pm$\,0.02} & $-$0.06   &         & $+$0.18    & $+$0.05  &          \\   
    GJ317         & {\bf $+$0.20\,$\pm$\,0.14} & $-$0.22   &         & $-$0.10    & $-$0.21  &          \\   
    GJ436         & {\bf $+$0.08\,$\pm$\,0.05} & $-$0.05   & $-$0.32 & $+$0.25    & $+$0.08  &  $+$0.00 \\   
    GJ581         & {\bf $-$0.15\,$\pm$\,0.03} & $-$0.25   & $-$0.33 & $-$0.10    & $-$0.21  &  $-$0.02 \\   
    GJ628         & {\bf $+$0.08\,$\pm$\,0.03} & $-$0.13   &         & $+$0.12\,c & $-$0.02  &          \\   
    GJ674         & {\bf $-$0.11\,$\pm$\,0.13} & $-$0.28   &         & $-$0.11    & $-$0.22  &          \\   
    GJ849 P82\&84 & {\bf $+$0.35\,$\pm$\,0.10} & $+$0.17   &         & $+$0.58    & $+$0.37  &  $+$0.49 \\   
    GJ876         & {\bf $+$0.12\,$\pm$\,0.15} & $+$0.02   & $-$0.12 & $+$0.37    & $+$0.24  &  $+$0.43 \\   
    \hline
   \end{tabular} 
   \label{tab:feh}
\end{table*}

For the metallicity determination we employed the method of fitting synthetic spectra to the observed spectra.
The analysis is based on the latest generation of MARCS model atmospheres
\citep{2008A&A...486..951G}.
These models give the temperature and pressure distribution in
radiative and hydrostatic equilibrium,
assuming radiative transport with mixing-length convection in a plane-parallel stellar atmosphere.
The formation of dust is not accounted for in the models, as it has been 
found to be less important in models of early-type M dwarfs
\citep[earlier than about M6 or \teff\ $\gtrsim$ 2600~K;][]{1997ApJ...480L..39J,2002ApJ...575..264T}.
Our sample includes rather early-type M dwarfs, see Table \ref{tab:targets}.

We used an improved version of the SME package \citep{1996A&AS..118..595V,2005ApJS..159..141V}.
This tool performs an automatic parameter optimization using a Levenberg-Marquardt chi-square minimization algorithm.
Synthetic spectra are calculated on the fly by a built-in spectrum synthesis code, for a set of global model parameters and specified spectral line data. Starting from user-provided initial values, synthetic spectra are computed for small offsets in different directions for a subset of parameters defined to be ``free''.
The required model atmospheres are interpolated in the grid of MARCS models available on the MARCS webpage\footnote{http://marcs.astro.uu.se}, using an accurate algorithm described in \citet[][Section 4.1]{2005ApJS..159..141V}.
Partial derivatives calculated from the corresponding parameter and chi-square values are used to approach the minimum in the chi-square surface.
In an independent step from the parameter optimization, the wavelength scale is corrected for any residual velocity shifts by a one-dimensional golden section search.
SME also has the option to apply a local fit of the continuum, but we did not use this functionality here. Continuum normalization was instead done during data reduction (see Section~\ref{sect:observations}).

A mask specifies the pixels in the observed spectrum which should be used to determine velocity 
corrections and to calculate the chi-square.
Mask definition is an important step in spectrum synthesis analysis.
The radial velocity correction was done in a first step, using most of the observed spectral region and 
solar metallicity synthetic spectra. This correction was applied to the observed spectra before defining 
the mask for the metallicity analysis. 
We placed the mask as consistently as possible for all programme stars to cover the maximum 
number of spectral lines not affected by blends between different species. Defects in the 
observed spectra caused by imperfect telluric correction or instrumental effects were masked 
out, as were the cores of strong lines. For some blended or contaminated lines, a part of the 
profile was included in the mask if considered useful for the fitting procedure.

\begin{figure*}[htbp]
  \centering  
   \resizebox{\hsize}{!}{\includegraphics{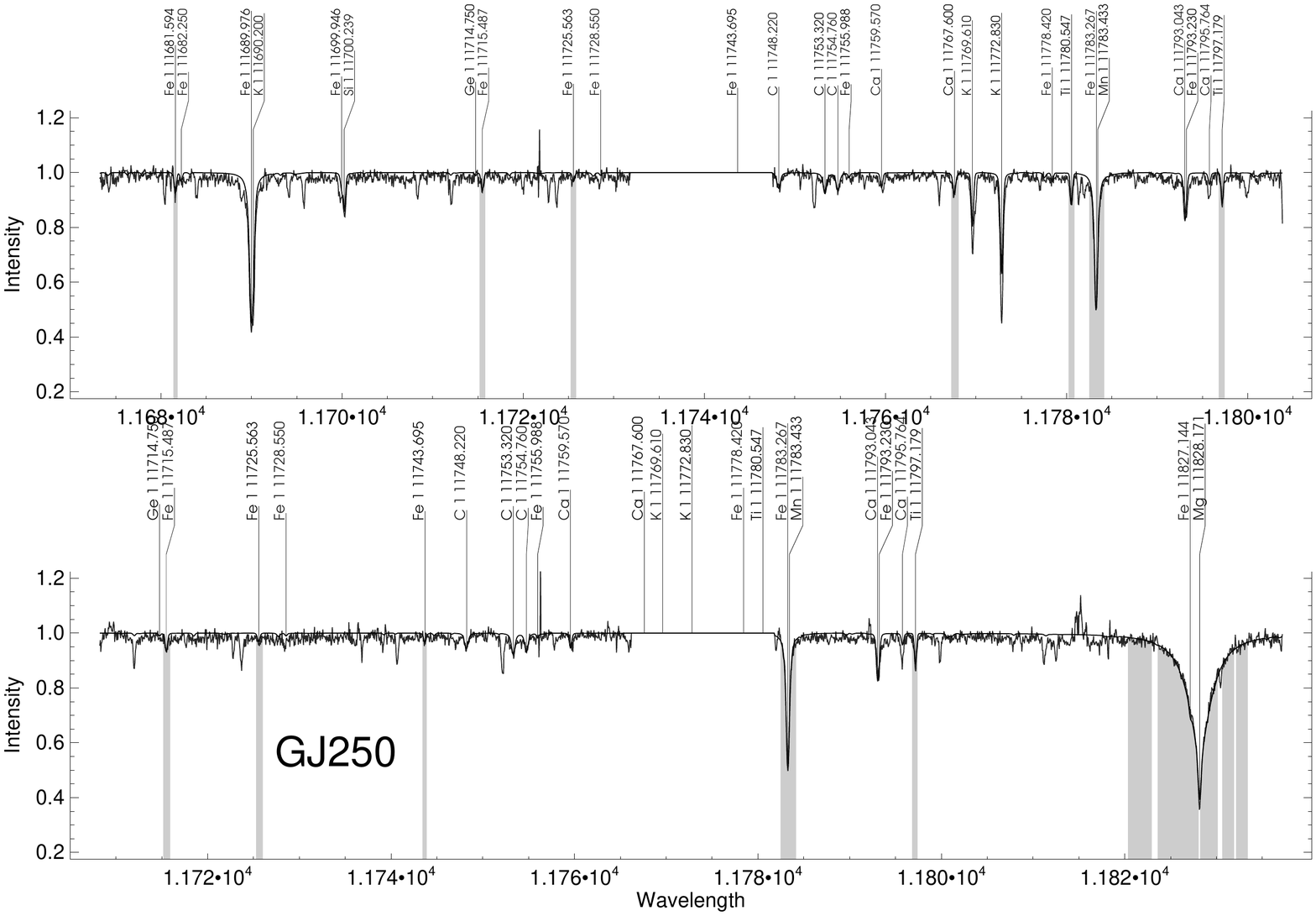}}
   \resizebox{\hsize}{!}{\includegraphics{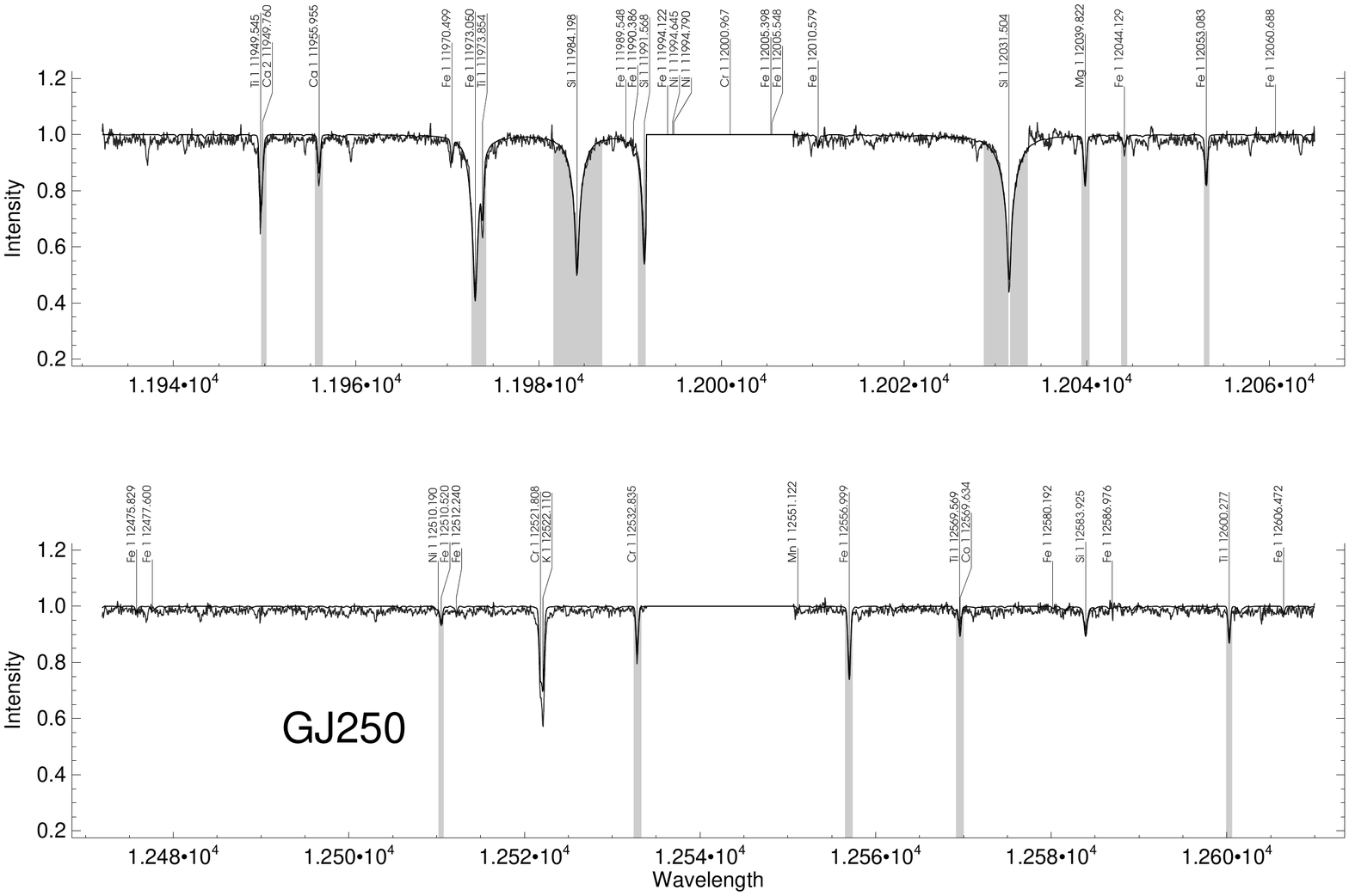}}
  \caption{The eight observed wavelength regions in period 82 plotted together with 
           the best fit synthetic spectrum for target GJ250A (primary in  
           double system). The mask used in the metallicity determination is marked
           with a grey shade.}
  \label{fig:GJ250A}
\end{figure*}

\begin{figure*}[htbp]
  \centering
   \resizebox{\hsize}{!}{\includegraphics{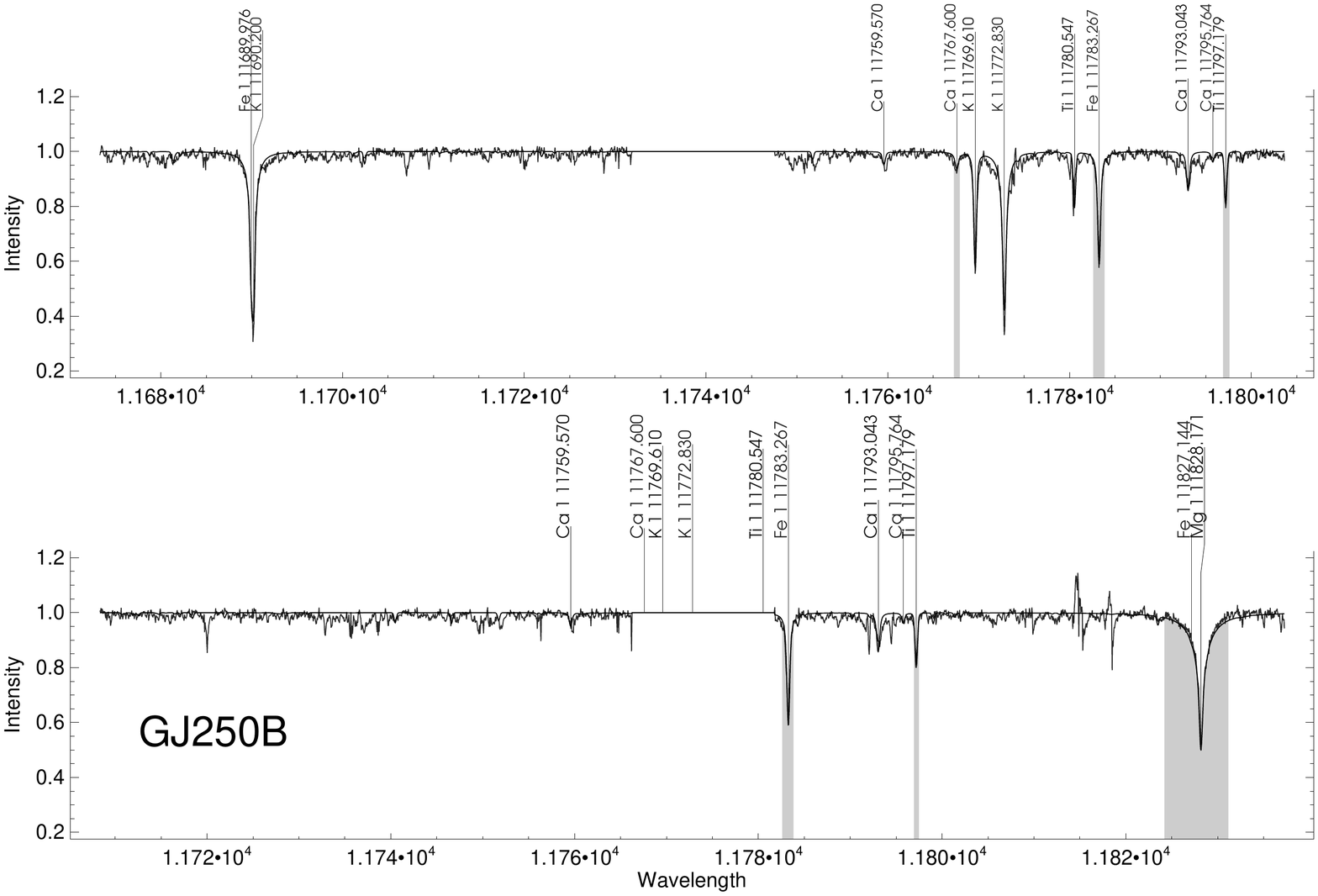}}
   \resizebox{\hsize}{!}{\includegraphics{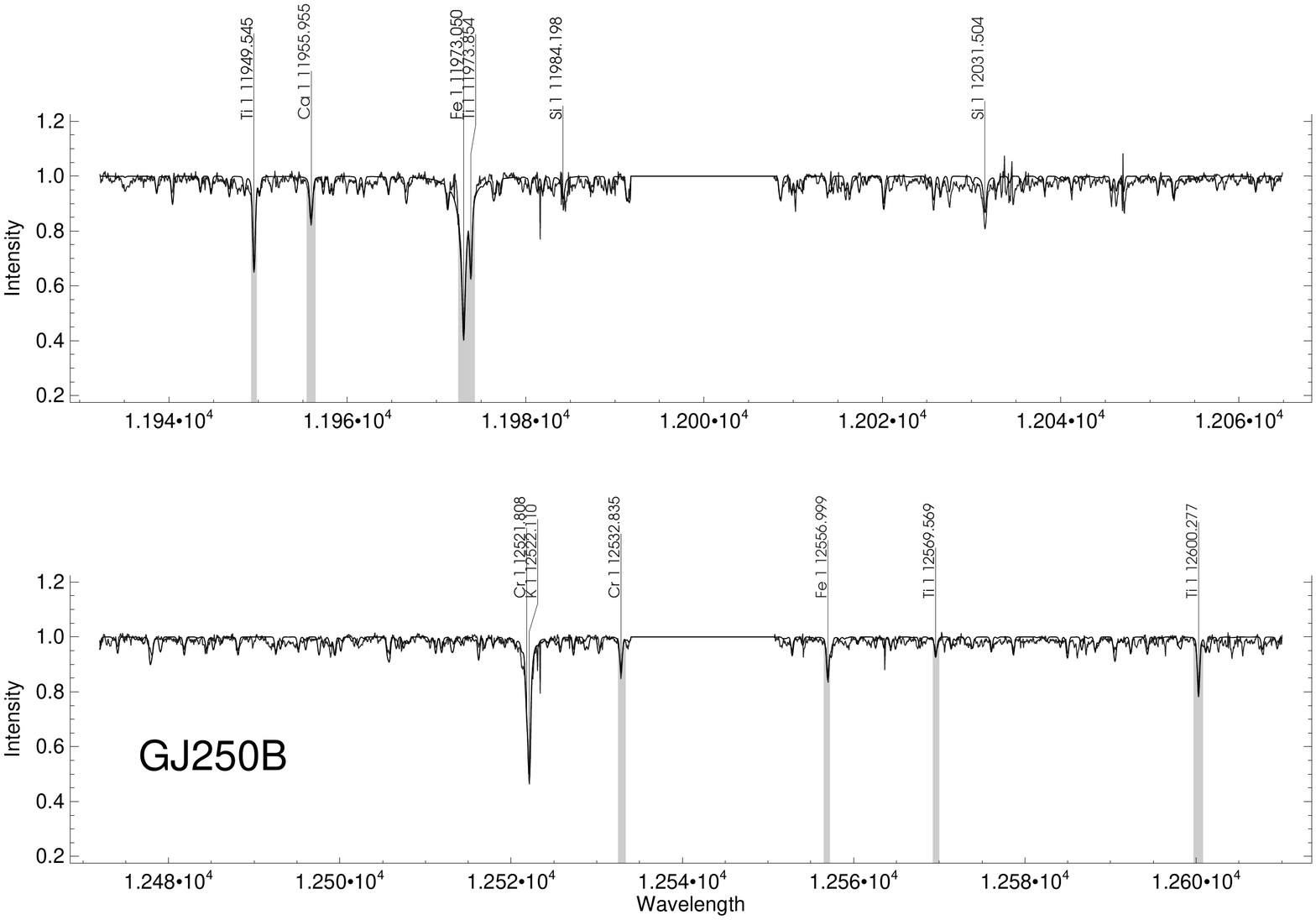}}
  \caption{The eight observed wavelength regions in period 82 plotted together with 
           the best fit synthetic spectrum for target GJ250B (secondary in  
           double system). The mask used in the metallicity determination is marked
           with a grey shade.}
  \label{fig:GJ250B}
\end{figure*}

The number of available lines in the telluric-free spectra used in the analysis was at maximum 30 
in a K dwarf and 23 in an M-dwarf spectrum.
The total number of lines included in the analysis of each target spectrum varied slightly due 
to differences in the data quality, imperfect telluric line removal and other non-physical 
spectral features. We note that the four potassium lines in our spectra
(1169.02, 1176.96, 1177.28, 1252.21\,nm) are most likely affected by 
non-LTE in the solar spectrum. This is suspected from the fact that when we adjusted the 
$gf$-values for these lines to fit a high-quality solar spectrum, the lines became too strong 
in the cooler, high-gravity M dwarfs. A discussion
on the solar non-LTE effect of these lines can be found in \citet{2006A&A...453..723Z},
where two of the K lines present in our spectra are explored. 
These authors derive a negative abundance correction for the K lines, meaning that the lines 
are stronger when calculated in non-LTE than for LTE.
Due to the higher densities, collisions may be expected to drive the atmospheres of M dwarfs 
towards LTE conditions. Although the non-LTE effects in the M dwarfs are probably smaller than 
in the Sun, they are as yet unknown to us. Hence, we decided not to use the K lines in the 
analysis and excluded them from our
line mask (see Figs.~\ref{fig:GJ250A} and \ref{fig:GJ250B}).

There are nine C~I lines apparent in the spectra of the Sun and the K dwarfs in our 
wavelength range. We tried to model these lines using atomic data from \citet{NIST10}, 
which are based on averaged calculated transition probabilities from two literature 
sources \citep[][using the ``velocity'' results of the latter]{4828TP,6070TP}, and 
van der Waals broadening parameters from \citet{BPM}.
However, the synthetic lines were too weak in the solar spectrum and at the same time 
somewhat too strong in the K-dwarf spectra, compared to the observations.  Hence, we 
were not able to derive ``astrophysical'' $gf$-values applicable to both types of stars. 
We suspect non-LTE effects to be the cause of this problem, which might be spectral-type 
dependent, in the sense that they are stronger in G-type dwarfs than in K-type dwarfs. 
This is supported by the fact that all of these lines are high-excitation lines, with 
lower level energies lying between 7.5 and 8.8~eV. Such levels are easily depopulated by 
photoionization, which leads to deviations from the LTE approximation.
We are not aware of any study of non-LTE effects for the carbon lines in question,
and did not include the carbon lines in the analysis.

For all other elements included in our investigation, we do not suspect any non-LTE effects 
based on the comparisons with the solar spectrum and the K-dwarf spectra. 
However, we cannot completely exclude any additional non-LTE effects based on our data and models. Unfortunately, 
non-LTE studies in the infrared region are rare, and investigations have focused on solar-type 
stars. 
The study by \citet{2004A&A...420..183A} indicates that lines 
of neutral Fe and Ca might be affected by departures from LTE in the coolest stars of their sample. 
However, the lowest \teff\ which they explore is close to 4500~K, and their spectral range extends 
from about 360~nm to 1~micron. For a review of non-LTE effects in optical spectra of FGK-type 
stars for a large number of elements see \citet{2005ARA&A..43..481A}.

The consistency of the SME solution with the selected surface gravity could be
tested using strong lines with well developed wings. While we are not sure about
non-LTE effects on potassium line strength we can try to use the shape of these lines for
a gravity test. In order to do that we solved for the surface gravity keeping all
the other parameters fixed from the optimal solution. The experiment was carried
out for one data set for two objects GJ 105B and GJ 628. We find that the surface 
gravities did not change by more than 0.03 dex confirming that the preset values are
consistent with the shapes of strong lines.

We let the overall metallicity, [Fe/H], and the macroturbulence parameter vary and solved
for both simultaneously. For the M dwarfs, the unknown microturbulence parameter was set to 
1\,kms$^{-1}$ and line broadening by rotation was neglected, as most of our M dwarfs are known to be 
relatively slow rotators (see Table~\ref{tab:paramM}). 
As macroturbulence was the only line-broadening parameter, which we included in the fit procedure, 
the value resulting in the best fit contains contributions from other broadening mechanisms, 
otherwise unaccounted for, e.g. variations in the instrumental profile, or rotational broadening. 
The derived values for our programme stars are below 1~kms$^{-1}$, except for GJ~876 (1.7~kms$^{-1}$), 
GJ~250A and B ($\approx$2~kms$^{-1}$), GJ~674 (3.8~kms$^{-1}$ comparable to its $v\sin i$ value in 
Table~\ref{tab:paramM}), and GJ~105B (4.6~kms$^{-1}$). 

To ensure that we end up in a global minimum when converging to a solution for the best fit,
we started from different initial values for the metallicity and plotted the resultant $\chi^2$ for each 
fit as a function of the determined metallicity. This also gave us an estimate of the 
uncertainties introduced by the fitting routine itself, and was included in the 
total error calculation. 
A majority of the lines included in the analysis are weak although some stronger lines
are present. We tested the convergence dependence on weak lines as compared to the wings of 
the stronger lines and found that the strong and weak lines contributed equally to the final $\chi^2$ 
solution. Both strong and weak lines resulted in equal metallicities.
We also estimated the Ca abundances for the M dwarfs observed in period 84, where
a number of Ca lines are present in the wavelength setup, and find [Ca/Fe] abundances of 0.0--0.1\,dex
with respect to solar values.

The total uncertainties were computed by perturbing
the atmospheric parameters by $\pm$100\,K in \teff\ and $\pm$0.1\,dex in \logg, and repeating the 
metallicity fit. The deviations with respect to the adopted [Fe/H] value were then added in 
quadrature to the uncertainties from the fitting routine itself.

\section{Results and discussion}
\label{sect:results}

We have carried out a careful analysis of high-resolution M and K dwarf 
spectra in the infrared J-band. To calibrate our metallicity scale, 
three binary systems consisting of a K-dwarf primary and an M-dwarf secondary were 
observed. The derived metallicities of the binary systems as well as the single M dwarfs
can be found in Table~\ref{tab:feh} and Figure \ref{fig:feh}. In this table and figure 
we also list metallicities determined by previous spectroscopic and photometric investigations.

\subsection{Binary systems}
Stars in binary systems have been formed out of the same molecular cloud
and are expected to have the same metallicities. For two of the three binary systems
present in this study, we derive very similar metallicities for both components.

The binary system {\bf GJ250 AB} consists of a K3-dwarf primary and an M2-dwarf 
secondary. The metallicity determinations for the two stars are 
consistent within the errors.
The synthetic spectra calculated for both stars show good agreement for both weaker
as well as stronger lines, including the potassium lines in the M dwarfs that are not 
included in the derivation of the best fit (see Figs.~\ref{fig:GJ250A} and 
\ref{fig:GJ250B}).

The M dwarf in the {\bf GJ105 AB} system shows too weak FeH lines
in comparison to the calculated spectrum based on the derived photometric 
effective temperature. We adjusted the temperature by +200\,K while
solving for [Fe/H] simultaneously as described in Section \ref{sect:atmparam}.
The difference in the determined metallicity of the components in the system 
was found to be 0.01\,dex, which is in agreement with coeval star formation.
We note, however, that the best fit synthetic spectrum does not give a satisfactory fit 
for all atomic lines as the stronger lines tend to be too broad for the established 
fit. This affect can arise from an incorrect surface gravity, although the
potassium line test described above does not support this explanation.

The third binary system, {\bf HD101930 AB} consists of an early K-dwarf
primary (K2) and an early M dwarf (M0-M1). The secondary is 
the only early M dwarf in our sample and we note that
the spectrum shows both features we recognise from the primary, such as strong Si lines,
and characteristics seen in the other M dwarfs, such as absent C lines and prominent 
FeH lines. In a careful study of the best fit synthetic spectrum
we noted that a majority of the lines fit rather well except for the strong
Si lines and the excluded K lines. We derived a metallicity of 0.09\,dex
which is 0.11\,dex lower than established for the primary of the system.
This difference is still on the order of the estimated uncertainties.

\subsection{Single M dwarfs}
\begin{figure}[htbp]
  \centering
    \includegraphics[scale=0.37]{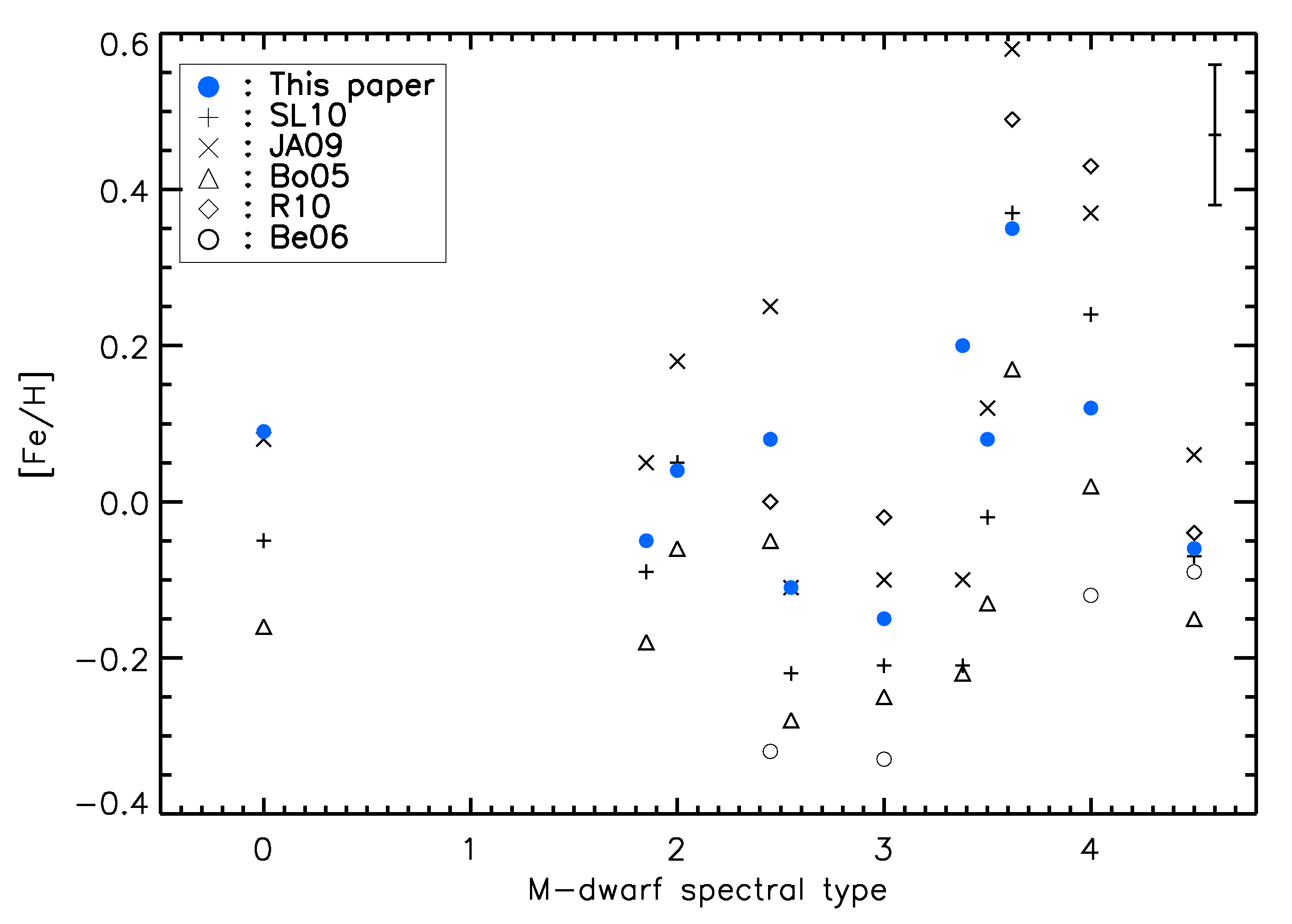}
  \caption{Derived metallicities for our sample in comparison with previous 
            abundance determinations (see Table \ref{tab:feh} for references) 
            as a function of spectral type (see Table \ref{tab:targets}).
            For an easier presentation the targets with equal spectral 
            types have been slightly shifted horizontally. In the upper 
            right corner we show an error bar indicating the mean error of the metallicities 
            derived in this paper.}
  \label{fig:feh}
\end{figure}

\begin{figure}[htbp]
  \centering
    \includegraphics[scale=0.37]{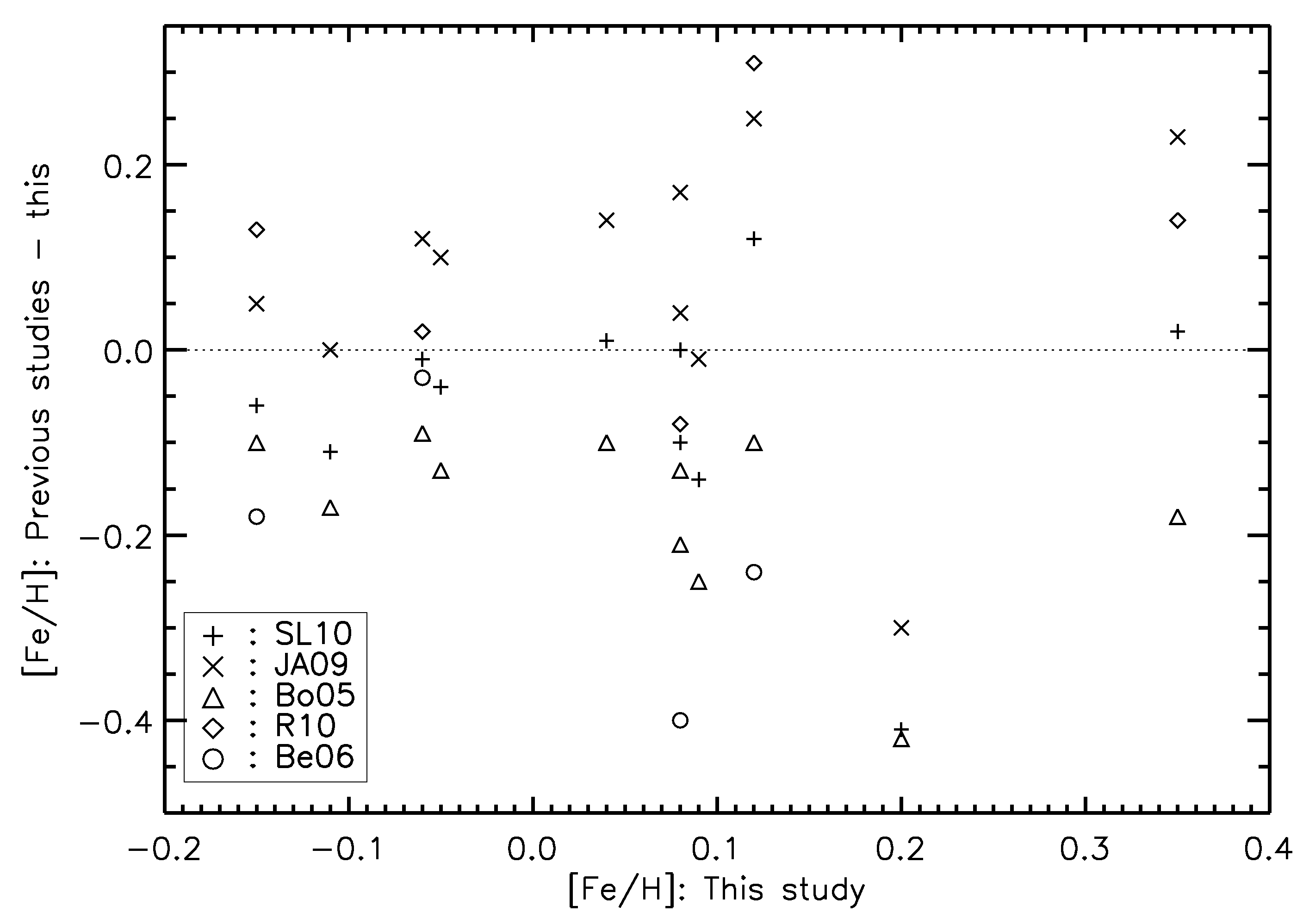} 
  \caption{Differences in metallicities for our sample in comparison with previous 
            abundance determinations (see Table \ref{tab:feh} for references) 
            as a function of metallicities from this study.
            A zero-line (dotted) to guide the eye is also plotted.}
  \label{fig:fehvsothers}
\end{figure}

For five of the eight single M dwarfs we derived 
metallicities for the first time based on high resolution
spectroscopy. For the three stars analysed in the optical 
by \citet{2006ApJ...653L..65B}, our metallicities are higher by $\gtrsim$0.2\,dex.
The recent study using spectroscopic indices by \citet{2010ApJ...720L.113R}, 
includes four of our targets and we derived lower metallicities for
three of these stars.
The photometric calibration by \citet{2005A_A...442..635B} returns
overall lower metallicities than we determined but the calibration
is claimed by \citet{2009ApJ...699..933J} to produce too low
[Fe/H] values. The latter authors corrected for this and we
do agree better with this higher metallicity scale. The best
agreement we seem to find with \citet{2010A&A...519A.105S}, 
who used a similar technique as \citet{2009ApJ...699..933J}.
After the completion of the present paper, we noted that \citet{2011arXiv1110.2694N} 
recently evaluated the photometric metallicity calibrations
of Bo05, JA10 and SL10, and rank the SL10 scale highest.
The metallicities derived in our work are shown in Figs. \ref{fig:feh} and \ref{fig:fehvsothers} together
with results from the discussed studies. The average differences
between our study and those of \citet{2010A&A...519A.105S}, 
\citet{2009ApJ...699..933J} and \citet{2005A_A...442..635B}
are 0.09, 0.13 and 0.17, respectively. 

\section{Conclusions}
\label{sect:conclusions}

We have derived a new metallicity scale based
on a careful spectroscopic analysis of 
high-resolution spectra (R$\sim$50,000) observed
in the J-band. Previous abundance studies are based on full 
spectroscopic analyses in the optical regime, low resolution 
spectroscopic infrared indices as well as purely photometric calibrations.
Observations in the infrared J-band have the advantage 
of few and weak molecular features (FeH) which 
allows for a precise continuum placement as compared to 
the optical wavelength regions where the continuum is heavily depressed
due to the many and strong TiO lines.
We find that we can correct for lines introduced by the Earth's 
atmosphere quite succesfully by using a rapidly rotating calibration 
star, and by applying a carefully determined continuum placement.
We verify the accuracy of the atmospheric models involved in
the metallicity determination by observing three binary systems that
are expected to have the same [Fe/H]. In two of the systems GJ105\,AB and GJ250\,AB
we find that the metallicities agree within 0.01 and 0.02\,dex, respectively. 
In the third binary system, HD101930\,AB, we find a discrepancy of 0.11\,dex,
which is consistent with the derived errors.

We test the convergence of the procedure by starting from 
different initial guesses and find consistent solutions although
some of the cooler objects show a greater spread in the
determined metallicity.

Our sample covers a restricted range in \teff\ (between $\approx$3200 and 3400\,K, 
and a single object at 3900\,K), and the extreme metallicities ($<-0.1$ and $>+0.2$) 
are not well covered. To explore the metallicity scale further, 
targets with a larger spread in the atmospheric parameters need to be 
observed, preferrably with an instrument that can cover a larger wavelength 
range efficiently.
Improving the completeness of molecular line lists will improve the
accuracy of the metallicity determinations, since
there are still a number of unknown molecular blends present.

We conclude that a high-resolution spectroscopic analysis in the near  
infrared is a reliable method for metallicity determinations in this
\teff\ regime. It is also the only method which will enable the determination 
of abundances of individual elements in M dwarfs.

\begin{acknowledgements}
We thank Jeff Valenti for developing SME, and for providing additional IDL routines
that were used in the analysis.
UH acknowledges support from the Swedish National Space Board (Rymdstyrelsen).
A\"O acknowledges support by V\"armlands nation (Uddeholms research scholarship).
This research has made use of the SIMBAD database, operated at CDS, Strasbourg, France.
NSO/Kitt Peak FTS data used here were produced by NSF/NOAO.
This publication makes use of data products from the Two Micron All
Sky Survey, which is a joint project of the University of Massachusetts
and the Infrared Processing and Analysis Center/California Institute of
Technology, funded by the National Aeronautics and Space Administration
and the National Science Foundation.
\end{acknowledgements}

\bibliographystyle{aa}
\bibliography{mdwarfs,lines,SpectralTypes,PreviousStudies}

\end{document}